# Decay Constants of Heavy-Light Mesons.

C.R. Allton[a] M. Crisafulli[a], V. Lubicz[a], G. Martinelli[a], G. Salina[b],
and
A. Bartoloni[a], C. Battista[a], S. Cabasino[a], N. Cabibbo[b], P. Marzano[a], P.S. Paolucci[a], J. Pech[a],
F. Rapuano[a], R. Sarno[a], G.M. Todesco[a], M. Torelli[a], W. Tross[a], P. Vicini[a]
*The Ape Collaboration*

[a] Dipartimento di Fisica, Universitá di Roma 'La Sapienza'
and INFN, Sezione di Roma, P.le A. Moro, I-00185 Rome, Italy.
[b] Dipartimento di Fisica, Universitá di Roma 'Tor Vergata'
and INFN, Sezione di Roma II, Via della Ricerca Scientifica 1, I-00173 Rome, Italy.

*The decay constants of the heavy-light pseudoscalar mesons are studied in a high statistics run using the Wilson action at $\beta = 6.0$ and $\beta = 6.2$, and the clover action at $\beta = 6.0$. The systematics of $O(a)$ discretisation errors are discussed. Our best estimates of the decay constants are: $f_D = 218(9)$ MeV, $f_D/f_{D_s} = 1.11(1)$ and we obtain preliminary values for $f_B$.*

## 1. INTRODUCTION

Lattice studies of the decay constants of heavy-light pseudoscalar mesons $f_P$ are precise only to the extent that they are not contaminated with discretisation errors due to the finiteness of the lattice spacing, $a$. If lattice QCD is to be predictive, it is essential that these systematic effects are understood and controlled. There are two approaches to reducing these errors: the first is to use the Wilson action, but at larger $\beta$ values where $a$ and hence these errors are smaller; the second is to use an 'improved formulation' such as the 'clover action' (see [1,2]) where the $O(a)$ errors have been systematically removed.

In this paper[1] we present results from around 100 configurations each for i) the Wilson action at $\beta = 6.0$, ii) the Wilson action at $\beta = 6.2$, and, iii) the clover action at $\beta = 6.0$. In analysing our results we are particularly careful to isolate the ground state to avoid higher state contamination. We also studied the dependence of the final results on the extrapolation in the heavy and light quark masses. We find that even with this relatively large statistics, it is difficult to discern the difference in $f_P$ between i) and ii). However, there is a clear difference between iii) and both the Wilson actions which we take as evidence for improvement.

All results presented were obtained on a 6.4 Gigaflop APE machine (see [4]). The full results of this work will be published elsewhere [5].

## 2. METHOD

Consider the correlation function

$$C_{L\Omega}(t) = \sum_{\vec{x}} < 0|A_0(\vec{x},t)\Omega^\dagger(\vec{0},0)|0 >, \qquad (1)$$

where $A_0(x) = \overline{Q}(x)\gamma_0\gamma_5 q(x)$, and $\Omega(x)$ is some pseudoscalar meson interpolating operator. To extract $f_P$ we also require

$$C_{\Omega\Omega}(t) = \sum_{\vec{x}} < 0|\Omega(\vec{x},t)\Omega^\dagger(\vec{0},0)|0 > . \qquad (2)$$

At large Euclidean times $C_{\Omega\Omega}(t)$ behaves as $\frac{Z_{\Omega\Omega}}{M} e^{-MT/2}\cosh(M(T/2-t))$ where $T$ is the temporal dimension of the lattice and $M$ is the mass of the pseudoscalar meson. We extract the raw lattice value of $f_P$ using the usual ratio method (see eg. [6]),

$$f_P^{lat} = \left\langle \frac{C_{L\Omega}(t)}{C_{\Omega\Omega}(t)} \coth(M(T/2-t)) \right\rangle \frac{\sqrt{Z_{\Omega\Omega}}}{M}, \qquad (3)$$

presented by C.R.Allton
[1] Other results for $f_P$ are presented in the proceedings [3].

2

Table 1
Simulation parameters and results obtained. The $f_B$ values are preliminary.

| | | | |
|---|---|---|---|
| $\beta$ | 6.0 | 6.2 | 6.0 |
| Action | Wilson | Wilson | clover |
| #Confs | 120 | 110 | 100 |
| Volume | $18^3 \times 64$ | $24^3 \times 64$ | $18^3 \times 64$ |
| $\kappa_q$ | 0.1530 | 0.1510 | 0.1425 |
| | 0.1540 | 0.1520 | 0.1432 |
| | 0.1550 | 0.1526 | 0.1440 |
| $\kappa_Q$ | 0.1255 | 0.1300 | 0.1150 |
| | 0.1320 | 0.1350 | 0.1200 |
| | 0.1385 | 0.1400 | 0.1250 |
| | 0.1420 | 0.1450 | 0.1330 |
| | 0.1455 | 0.1500 | |
| $\Omega$ | $P_5$ Local | $P_5$ Local | $P_5$ Unrotated |
| $t_{fit}$ | 15-28 | 20-28 | 15-28 |
| $\sigma a^2$ [7] | 0.050 | 0.0271 | 0.050 |
| $f_D/f_\pi$ | 1.18(6) | 1.32(10) | 1.65(7) |
| using eq.(5) | 1.61(7) | 1.67(13) | |
| $f_{D_s}/f_\pi$ | 1.32(5) | 1.53(10) | 1.82(7) |
| using eq.(5) | 1.84(7) | 1.96(13) | |
| $f_B$[MeV] | 112(8) | 120(11) | 197(18) |
| using eq.(5) | 191(11) | 176(17) | |

where $< \ldots >$ is a weighted average over time. The physical value of $f_P$ is simply

$$f_P \equiv \frac{<0|A_0|P(\vec{q}=\vec{0})>}{M} = f_P^{lat}\, Z_{ren}(a)\, a^{-1}, \quad (4)$$

where $Z_{ren}$ is the renormalisation constant of the axial current.

The aim of this work is to uncover the dependence of $f_P^{lat}$ on $a$. Hence we choose to study the ratio $f_P/f_\pi$. This eliminates $Z_{ren}$ which has sizeable uncertainties.

The parameters used in the simulations are listed in table 1. The two different volumes are chosen to ensure that the physical dimensions of the lattice are approximately the same at the two $\beta$ values. The $\kappa_Q$ values correspond to quarks of up to around the charm mass. In the table the type of interpolating operator $\Omega$ is also listed. Also in the table the fitting window for the correlation function (see eq.(3)) is displayed. This range is chosen by demanding that the contamination of the excited states in the parameters of the fits (ie. $< \frac{C_{\Omega\Omega}}{C_{\Omega\Omega}} \coth >$, M and $Z_{\Omega\Omega}$) is at most 20% of the statistical errors. We impose this strict criteria to stop the systematic effects of the higher states swamping the $a$ dependence of $f_P^{lat}$. (This is illustrated in detail elsewhere [5].)

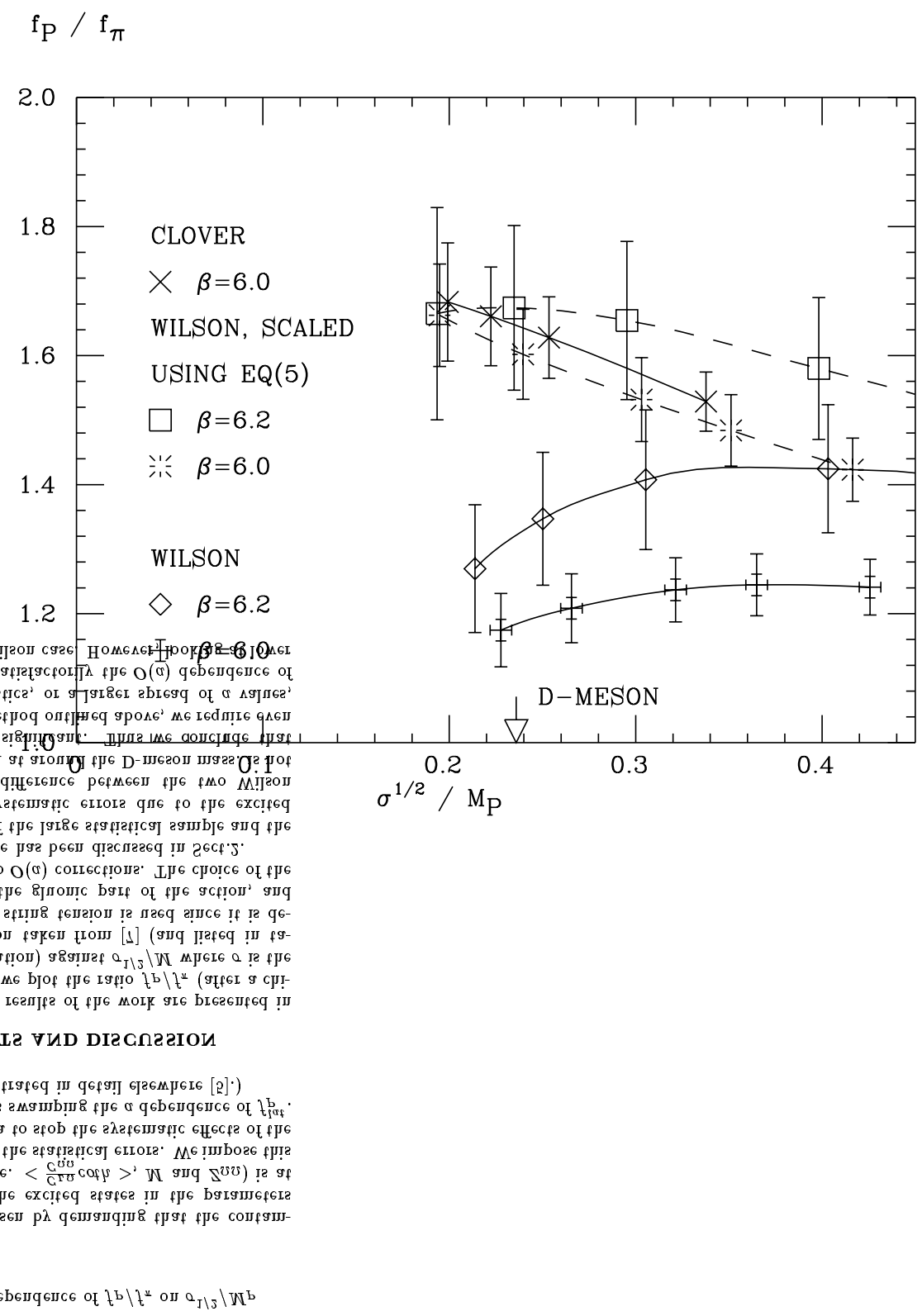

Figure 1. Dependence of $f_P/f_\pi$ on $\sigma^{1/2}/M_P$

## 3. RESULTS AND DISCUSSION

The main results of the work are presented in Fig.1. Here we plot the ratio $f_P/f_\pi$ (after a chiral extrapolation) against $\sigma^{1/2}/M$ where $\sigma$ is the string tension taken from [7] (and listed in table 1). The string tension is used since it is derived from the gluonic part of the action, and hence has no O($a$) corrections. The choice of the $y$−coordinate has been discussed in Sect.2.

In spite of the large statistical sample and the negligible systematic errors due to the excited states, the difference between the two Wilson data in Fig.1 at around the D-meson mass is not statistically significant. Thus we conclude that using the method outlined above, we require even higher statistics, or a larger spread of $a$ values, to uncover satisfactorily the O($a$) dependence of $f_D$ in the Wilson case. However, looking at lower mass values, it is tempting to say that the $\beta = 6.2$ values are higher than those of $\beta = 6.0$. Turning to the clover data, we see it is clearly larger than the Wilson data, particularly around the D-meson mass, which confirm the observations of UKQCD [8]. Note also, that the theoretical expectation is that $f_P/f_\pi$ increases with decreasing $a$. These remarks imply that the approach to the continuum limit appears considerably more rapid in the case of the clover compared with the Wilson actions.

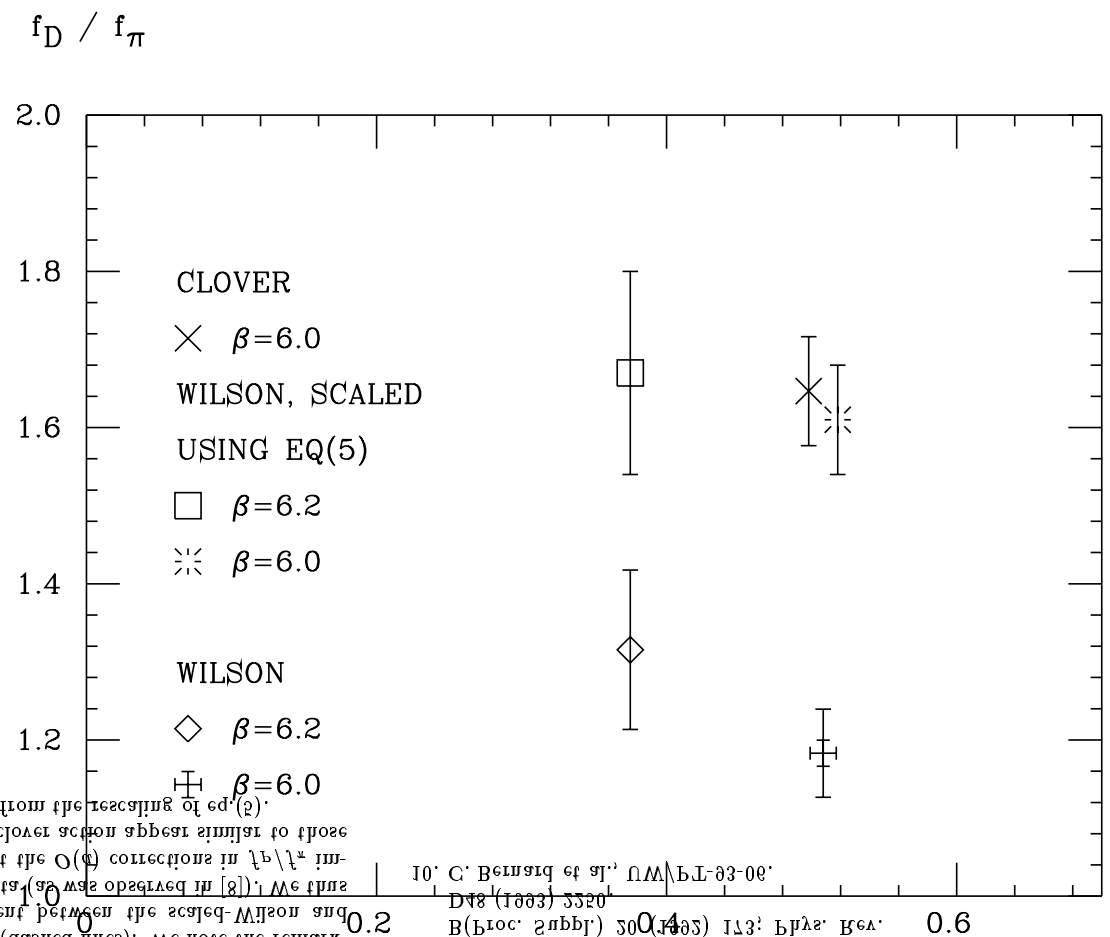

Figure 2. $f_D/f_\pi$ against $a$

Recently a suggestion has been proposed to try to correct some of the $O(a)$ effects apparent in the Wilson action [9,10] In this approach, the quark fields $Q, q$ and the mass $M$ are rescaled by

$$q \rightarrow \mathcal{N}(\kappa)\, q, \qquad Ma \rightarrow Ma + \tilde{m}_2 - \tilde{m} \qquad (5)$$

where $\mathcal{N}(\kappa) = (\frac{4\kappa_{crit}}{\kappa} - 3)^{1/2}$, $\tilde{m} = log(\mathcal{N}(\kappa))$ and $\tilde{m}_2 = e^{\tilde{m}} sinh\, \tilde{m}/(sinh\, \tilde{m} + 1)$.

Fig.1 shows the results of this rescaling on the Wilson data (dashed lines). We note the remarkable agreement between the scaled-Wilson and the clover data (as was observed in [8]). We thus conclude that the $O(a)$ corrections in $f_P/f_\pi$ implicit in the clover action appear similar to those which result from the rescaling of eq.(5).

Finally, by interpolating to the D-meson mass one obtains the results for $f_D/f_\pi$ listed in table 1 and plotted as a function of $a$ in Fig.2. Note that the two scaled-Wilson and the clover results are totally compatible. However it is clear that the size of the errors in the $\beta = 6.2$ Wilson data point preclude a meaningful extrapolation to the continuum limit and therefore an estimate of the systematic errors due to the finiteness of $a$ is not possible with the present statistics. We therefore quote our clover result as our best estimate of $f_D/f_\pi$ and hence obtain $f_D = 218(9)MeV$ where the error is purely statistical.

By fixing the $\kappa$ value corresponding to the strange quark using the mass of the K-meson (and the scale from the string tension), we obtain the values for $f_{D_s}/f_\pi$ shown in the table. For the same arguments as in the $f_D/f_\pi$ case, we use the clover result to quote $f_{D_s} = 240(9)MeV$, or alternatively $f_{D_s}/f_D = 1.11(1)$.

In the table we list the values of $f_B$ obtained for each simulation using the expected heavy quark scaling of $f_P\sqrt{M} \approx \Phi_{inf} + \frac{\Phi'}{M} + \frac{\Phi''}{M^2}$. These results are preliminary since we are still investigating the scaling behaviour of $f_B$ with $M$.